
%
\input phyzzx
\input tables
\catcode`@=11
%
%
\newtoks\KUNS
\newtoks\HETH
\newtoks\monthyear
\Pubnum={KUNS~\the\KUNS\cr HE(TH)~\the\HETH}
\KUNS={1252}
\HETH={94/04}
\monthyear={March, 1994}
\def\p@bblock{\begingroup \tabskip=\hsize minus \hsize
    \baselineskip=1.5\ht\strutbox \topspace-2\baselineskip
    \halign to\hsize{\strut ##\hfil\tabskip=0pt\crcr
    \the\Pubnum\cr hep-th/9403284\cr \the\monthyear\cr }\endgroup}
\def\bftitlestyle#1{\par\begingroup \titleparagraphs
    \iftwelv@\fourteenpoint\else\twelvepoint\fi
    \noindent {\bf #1}\par\endgroup}
\def\title#1{\vskip\frontpageskip \bftitlestyle{#1} \vskip\headskip}
%
%
\def\acknowledge{\par\penalty-100\medskip \spacecheck\sectionminspace
    \line{\hfil ACKNOWLEDGEMENTS\hfil}\nobreak\vskip\headskip}
%
%

%
\def\journal#1&#2(#3){\begingroup \let\journal=\dummyj@urnal
    \unskip, \sl #1\unskip~\bf\ignorespaces #2\rm
    (\afterassignment\j@ur \count255=#3) \endgroup\ignorespaces}
\def\andjournal#1&#2(#3){\begingroup \let\journal=\dummyj@urnal
    \sl #1\unskip~\bf\ignorespaces #2\rm
    (\afterassignment\j@ur \count255=#3) \endgroup\ignorespaces}
\def\andvol&#1(#2){\begingroup \let\journal=\dummyj@urnal
    \bf\ignorespaces #1\rm
    (\afterassignment\j@ur \count255=#2) \endgroup\ignorespaces}
\def\MPL{Mod.~Phys.~Lett.}
\def\NP{Nucl.~Phys.}
\def\PL{Phys.~Lett.}
\def\PR{Phys.~Rev.}
\def\PRL{Phys.~Rev.~Lett.}
\def\PTP{Prog.~Theor.~Phys.}
\def\ZP{Z.~Phys.}
\catcode`@=12
%

\titlepage

\title{Hierarchical Mass Matrices
       \break in a Minimal SO(10) Grand Unification I}

\author{Masako Bando\rlap,
\foot{On leave of absence from
{\it Aichi University, Aichi 470-02, Japan}.}
Izawa K.-I., and Tomohiko Takahashi}
\address{Department of Physics, Kyoto University \break
                    Kyoto 606, Japan}

\abstract{
We consider a minimal SO(10) unified model
with horizontal Peccei-Quinn symmetry.
The hierarchical structure of quark-lepton mass matrices
is naturally implemented by the remnants
of certain irrelevant terms.
Georgi-Jarlskog relations are also realized
due to the horizontal symmetry.
}

\endpage

\doublespace


\def\c{\varepsilon}

\def\f{\phi}

\def\h{\theta}

\def\l{\lambda}
\def\m{\mu}
\def\n{\nu}
\def\p{\psi}

\def\t{\tau}

\def\w{\omega}

\def\D{\Delta}

\def\F{\Phi}

\def\S{\Sigma}

\def\H{${\rm U}(1)_{\rm H}$}
\def\U{{\rm SO}(10)}
\def\PS{{\rm SU}(2)_{\rm L} \times {\rm SU}(2)_{\rm R}
        \times {\rm SU}(4)_{\rm C}}
\def\SM{{\rm U}(1)_{\rm Y} \times {\rm SU}(2)_{\rm L}
        \times {\rm SU}(3)_{\rm C}}

\def\o{\overline}


\REF\Suz{For a review, A.~Suzuki, KEK Preprint 93-96.}

\REF\Eli{J.~Ellis and M.K.~Gaillard \journal \PL &B88 (79) 315;
         \nextline
         R.~Barbieri, J.~Ellis, and M.K.~Gaillard
         \journal \PL &B90 (80) 249; \nextline
         J.~Ellis, D.V.~Nanopoulos, and K.~Tamvakis
         \journal \PL &B121 (83) 123.}

\REF\Nan{D.V.~Nanopoulos and M.~Srednicki \journal \PL &B124 (83) 37;
         \nextline
         M.~Bando, Y.~Sato, and S.~Uehara \journal \ZP &C22 (84) 251;
         \nextline
         G.~Anderson, S.~Raby, S.~Dimopoulos, L.J.~Hall,
         and G.D.~Starkman, preprint LBL-33531.}

\REF\Dav{A.~Davidson and K.C.~Wali \journal \PRL &48 (82) 11;
         \nextline
         A.~Davidson, V.P.~Nair, and K.C.~Wali
         \journal \PR &D29 (84) 1504; 1513; \nextline
         S.~Dimopoulos \journal \PL &B129 (83) 417; \nextline
         S.~Dimopoulos and H.M.~Georgi \journal \PL &B140 (84) 67;
         \nextline
         J.~Bagger, S.~Dimopoulos, H.~Georgi, and S.~Raby,
         in {\sl Fifth Workshop on Grand Unification},
         ed. K.~Kang {\it et al.} (World Scientific, 1984); \nextline
         A.S.~Joshipura \journal \ZP &C38 (88) 479; \nextline
         G.~Lazarides and Q.~Shafi \journal \NP &B350 (91) 179;
         \andvol &B364 (91) 3.}

\REF\Giu{G.F.~Giudice \journal \MPL &A7 (92) 2429.}

\REF\Sla{R.~Slansky \journal Phys.~Rep. &79 (81) 1.}

\REF\Din{For a review, M.~Dine, preprint SCIPP 93/31.}

\REF\Sik{P.~Sikivie \journal \PRL &48 (82) 1156; \nextline
         R.N.~Mohapatra and G.~Senjanovi{\' c}
         \journal \ZP &C20 (83) 365; \nextline
         R.N.~Mohapatra, S.~Ouvry, and G.~Senjanovi{\' c}
         \journal \PL &B126 (83) 329; \nextline
         D.~Chang and R.N.~Mohapatra \journal \PR &D32 (85) 293.}

\REF\Laz{G.~Lazarides, Q.~Shafi, and C.~Wetterich
         \journal \NP &B181 (81) 287; \nextline
         K.S.~Babu and R.N.~Mohapatra \journal \PRL &70 (93) 2845.}

\REF\Geo{H.~Georgi and C.~Jarlskog \journal \PL &B86 (79) 297;
         \nextline
         H.~Georgi and D.V.~Nanopoulos \journal \NP &B159 (79) 16;
         \nextline J.~Harvey, P.~Ramond, and D.B.~Reiss
         \journal \NP &B199 (82) 223.}

\REF\Bur{A.J.~Buras, J.~Ellis, M.K.~Gaillard, and D.V.~Nanopoulos
         \journal \nextline \NP &B135 (78) 66; \nextline
         D.V.~Nanopoulos and D.A.~Ross \journal \NP &B157 (79) 273.}

\REF\Wil{F.~Wilczek and Z.~Zee \journal \PL &B70 (77) 418;
         \andjournal \PRL &42 (79) 421; \nextline
         H.~Fritzsch \journal \PL &B70 (77) 436; \andvol &B73 (78) 317.}

\REF\Des{N.G.~Deshpande, E.~Keith, and P.B.~Pal
         \journal \PR &D46 (92) 2261; \nextline
         R.N.~Mohapatra and M.K.~Parida \journal \PR &D47 (93) 264;
         \nextline
         L.~Lavoura and L.~Wolfenstein \journal \PR &D48 (93) 264;
         \nextline
         A.~Datta, S.~Pakvasa, and U.~Sarkar
         \journal \PL &B313 (93) 83.}

\REF\Del{H.~Georgi and D.V.~Nanopoulos \journal \PL &B82 (79) 95
         \nextline
         F.~del Aguila and L.E.~Ib{\' a}{\~ n}ez
         \journal \NP &B177 (81) 60; \nextline
         R.N.~Mohapatra and G.~Senjanovi{\' c}
         \journal \PR &D27 (83) 1601; \nextline
         H.E.~Haber and Y.~Nir \journal \NP &B335 (90) 363.}

\REF\Par{Particle Data Group \journal \PR &D45 (92) S1.}

\REF\Fuk{For a review, M.~Fukugita and T.~Yanagida,
         preprint YITP/K-1050.}

\REF\Sak{R.N.~Mohapatra and B.~Sakita \journal \PR &D21 (80) 1062;
         \nextline
         F.~Wilczek and Z.~Zee \journal \PR &D25 (82) 553; \nextline
         K.~Itoh, T.~Kugo, and H.~Kunitomo \journal \PTP &75 (86) 386.}

\sequentialequations

%
{\caps 1. Introduction}

Grand unification is a promising approach to
fundamental theory of elementary particles.
It generically predicts decay of the proton into leptons.
Long life of the proton
\refmark{\Suz}
implies that the supposed
grand unification scale $M_U$ is close
to the Planck scale $M_P \sim 10^{19}$ GeV,
which seems a plausible cut-off scale
for field-theoretical models of fundamental interactions.
Hence, at the unification scale, so-called irrelevant
(or nonrenormalizable)
terms may play a significant role in unified-model building.

When we come to the weak scale, most of the irrelevant terms are
indeed irrelevant, at least, perturbatively.
However, Yukawa interactions involving Higgs fields
with the unification-scale
vacuum expectation value (VEV)
exhibit sizable remnant effects
even at the weak scale\rlap.
\refmark{\Eli}
These effects provide a small quantity $M_U/M_P$,
which is expected to be a seed for the hierarchical structure
of the fermion mass matrices\rlap.
\refmark{\Nan}

In this paper,
we consider a minimal \U\ unified model
with horizontal Peccei-Quinn symmetry
\refmark{\Dav}
as an attempt to realize the above scenario:
We demand that the coupling constants in the theory
be of order unity at the cut-off scale,
and postulate necessary gauge hierarchy.
Then the hierarchical structure of mass matrices
is naturally implemented by the
remnant effects
in combination with the horizontal symmetry,
which simultaneously lead to a texture
{\` a} la Giudice\rlap.
\refmark{\Giu}

%
{\caps 2. The Model}

In this section, we expose a minimal \U\  grand unified model
with horizontal Peccei-Quinn symmetry \H.

Let us first present
the field contents of the model.
We have:
three left-handed Weyl spinors $\p_1, \p_2, \p_3$
corresponding to three generations of fermions;
three complex scalar multiplets $H$, ${\bar \D}$, $\F$ which cause
spontaneous symmetry breaking at the weak, intermediate,
and unification scales, respectively;
along with the gauge field for the gauge group \U.
Their \U\ and \H\ representations are shown in table 1.

\begintable
Field |\U      |\H||Field       |\U            |\H    \crthick
$\p_1$|$\bf 16$|9 ||$H$         |$\bf 10$      |$-2$  \cr
$\p_2$|$\bf 16$|5 ||${\bar \D}$ |$\o {\bf 126}$|$-10$ \cr
$\p_3$|$\bf 16$|1 ||$\F$        |$\bf 210$     |$-8$  \endtable

\centerline{
Table 1: Representations of the field multiplets.}

\nextline
\noindent
Note that the \H\  charges of the scalars are given by
$(-2)$ times the tensorial degrees of the corresponding \U\
representations.
Namely, the \H\  charge of the \U\  vector $H_j$
is $(-2) \times 1 = -2$,
that of the antisymmetric tensor of degree five
${\bar \D}_{jklmn}$ is $(-2) \times 5 = -10$, and
that of the antisymmetric tensor of degree four
$\F_{jklm}$ is $(-2) \times 4 = -8$.
We restate the point that
this \H\ symmetry plays a dual role to implement
both hierarchy and pattern of mass matrices
which lead to realistic quark-lepton masses and mixings.

We next exhibit the sequence of gauge symmetry breaking
in the model.
The unification group \U\ is broken to the standard-model group
$\SM$ via the Pati-Salam group $\PS$:
$$
 \eqalign{
  \U &\buildrel \VEV{\F} = M_U \over
       {\hbox to 40pt{\rightarrowfill}} \PS \cr
     &\buildrel \langle {\bar \D} \rangle = M_I \over
       {\hbox to 40pt{\rightarrowfill}} \SM
      \buildrel \VEV{H} = M_W \over
       {\hbox to 40pt{\rightarrowfill}}
        {\rm U}(1)_{\rm EM} \times {\rm SU}(3)_{\rm C}. \cr
 }
 \eqn\SEQ
$$
The unification-scale breaking is induced
by condensation of the $(1, 1, 1)_{\bf 210}$
component in the Higgs field $\F$,
the intermediate one
by that of the $(1, 3, {\o {10}})_{\o {\bf 126}}$
component
in the Higgs field ${\bar \D}$,
and the electroweak one by that of the $(2, 2, 1)_{\bf 10}$
component in the Higgs field $H$.
More precisely, the $(2, 2, 15)_{\o {\bf 126}}$ component
also contributes to the electroweak breaking, which
will be the subject of section 4.
Here we have denoted the irreducible decompositions
of \U\ representations
\refmark{\Sla}
under the Pati-Salam group as follows:
$$
 \eqalign{
  {\bf 10} &= (2, 2, 1) + (1, 1, 6), \cr
  {\o {\bf 126}} &= (1, 1, 6) + (3, 1, 10) + (1, 3, {\o {10}})
                                           + (2, 2, 15), \cr
  {\bf 210} &= (1, 1, 1) + (1, 1, 15) + (2, 2, 6) \cr
  &\quad + (3, 1, 15) + (1, 3, 15) + (2, 2, 10) + (2, 2, {\o {10}}). \cr
 }
 \eqn\DECOM
$$
We note that the VEV of the Higgs field increases
with the dimension of its representation.

A comment is in order about the horizontal
Peccei-Quinn symmetry \H,
which is broken at the unification scale
since the Higgs field $\F$ has non-zero \H\ charge.
This results in a superlight axion, which seems inappropriate
if cosmological constraints are taken seriously\rlap.
\refmark{\Din}
We then introduce soft-breaking terms for the \H\ symmetry\rlap,
\refmark{\Sik}
which eliminate the troublesome axion.

%
{\caps 3. Yukawa Interactions}

Now we consider Yukawa interactions
at and above the unification scale.
The \U\ representation of the fermions obeys
a composition law
$$
  {\bf 16} \times {\bf 16} = {\bf 10_s}
   + {\bf 120_a} + {\bf 126_s},
 \eqn\COMP
$$
where
{\bf 120} is the \U\ antisymmetric tensor of degree three, and
the indices {\bf s} and {\bf a} show that the corresponding
representations consist of only
symmetric and antisymmetric combinations in generation indices,
respectively.

Let $y$'s and $z$'s be dimensionless coupling constants of order one.
Owing to the \U\ $\times$ \H\ symmetry,
dimension-four terms which contribute to the mass
matrices are given by
$$
  {\cal O}_4 = y_{33} \p_3 \p_3 H
   + z_{22} \p_2 \p_2 {\bar \D}
    + z_{13} (\p_1 \p_3 + \p_3 \p_1) {\bar \D},
 \eqn\FOUR
$$
with their hermitian conjugates added.
When necessary,
this addition of hermitian conjugates should
be understood also in the following.

Under the circumstances that
the $(1, 1, 1)_{\bf 210}$ component in the Higgs field $\F$
develops the unification-scale VEV,
higher-dimensional terms written below
also contribute
to the mass matrices through their remnants
(See Appendix A.1):
Dimension-five terms are given by
$$
 \eqalign{
  {\cal O}_5 &= y_{13} (\p_1 \p_3 - \p_3 \p_1) H_j \F_{jklm}
   + y_{23} (\p_2 \p_3 - \p_3 \p_2) H^*_j \F_{jklm} \cr
  &\quad + {1 \over 2}z_{11} \p_1 \p_1 {\bar \D}_{jklmn} \F_{mnpq}
   + {1 \over 2}z_{33} \p_3 \p_3 {\bar \D}_{jklmn} \F^*_{mnpq}, \cr
 }
 \eqn\FIVE
$$
where * denotes complex conjugation.
Dimension-six ones by
$$
  {\cal O}_6 = {1 \over 3!}y_{11} \p_1 \p_1 H_j \F_{jklm} \F_{klmn}
   + {1 \over 3!}y_{12} (\p_1 \p_2 + \p_2 \p_1)
    H^*_j \F_{jklm} \F_{klmn},
 \eqn\SIX
$$
where we have omitted terms such as $\p_3 \p_3 H \F^* \F$,
which gives negligible corrections
to the leading term $\p_3 \p_3 H$
when $M_U/M_P \ll 1$.

The above terms of dimensions less than seven exhaust the leading
sources of the mass matrices that respect the symmetry \U\ $\times$ \H.
They are combined into
a Lagrangian density
$$
  {\cal L}_Y = -{\cal O}_4 - {1 \over M_P}{\cal O}_5
              - {1 \over M_P^2}{\cal O}_6,
 \eqn\LAGL
$$
with the Planck scale $M_P$ as the cut-off in the theory.
We note that the presence of $H^*$ coupling terms in \FIVE\ and \SIX\
is important for realizing sizable mixings among fermions,
which will be explored in section 5.

%
{\caps 4. Induced VEV}

Before proceeding to work out the mass matrices
in the standard-model framework,
we need some investigation on induced
VEVs which affect them.

We have assumed the gauge hierarchy
that the Higgs components
$(1, 1, 1)_{\bf 210}$,
\break
$(1, 3, {\o {10}})_{\o {\bf 126}}$,
and $(2, 2, 1)_{\bf 10}$ acquire appropriate VEVs.
Since the net unbroken symmetry
is U$(1)_{\rm EM} \times$ SU$(3)_{\rm C}$,
other components depicted in \DECOM\ also get non-zero VEVs
which respect the unbroken symmetry.
In particular, the $(2, 2, 15)_{\o {\bf 126}}$ component
may have VEVs which directly make sizable
contribution to the mass matrices\rlap.
\refmark{\Laz}

The Higgs potential for the $(2, 2, 15)_{\o {\bf 126}}$
component contains terms of the form
$$
  {\l \over M_P} \S^* \VEV{\D}\VEV{\bar \D}\VEV{H}\VEV{\F},
 \eqn\HPOT
$$
where $\S^*$ denotes
fields in the $(2, 2, 15)_{\o {\bf 126}}$ component
with masses of order $M_\S$,
$\l$ denotes coupling constants of order one,
and $\D = {\bar \D}^*$.
These terms give leading contributions to the effective potential,
which implies that the fields
$\S$ possess VEVs
$$
  \VEV{\S} \sim \c \l {M_I^2 \over M_\S^2} \VEV{H},
 \eqn\INVEV
$$
where $\c = M_U / M_P$.

Let us put a crucial assumption
$M_\S \sim M_I$.
Then the expression \INVEV\ indicates
$$
  \VEV{\S} \sim \c \VEV{H}.
 \eqn\OINVEV
$$
This suggests that the $(2, 2, 15)_{\o {\bf 126}}$ component
makes small contribution to the electroweak breaking,
which is of order $\c$ relative to the weak scale.

%
{\caps 5. Mass Matrices}

This section is devoted to derivation of mass matrices in the
standard-model framework from the Yukawa interactions \LAGL\
presented in section 3.

We write the VEVs of the fields $H$ and $\S$
as $v_{t, b}$ and $w_{c, s}$, respectively.
Here the indices $t, b$ or $c, s$
distinguish the two
VEVs which the corresponding field $H$ or $\S$
possesses (See Appendix A.1).
Note that $w_{c, s} \sim \c v_{t, b}$,
as explained in the previous section.

Analyses based on the \U\ representation contents
in the terms \LAGL\ reveal the leading entries of
quark-lepton mass matrices as follows:
\nextline
$i)$ up-quark mass matrix
$$
 \eqalign{
  M_u &= v_t \pmatrix{-\c^2 y_{11} & 0           & \c y_{13}  \cr
                      0            & 0           & 0          \cr
                      -\c y_{13}   & 0           & y_{33}     \cr}
     + v^*_b \pmatrix{0            & \c^2 y_{12} & 0          \cr
                      \c^2 y_{12}  & 0           & -\c y_{23} \cr
                      0            & \c y_{23}   & 0          \cr} \cr
      &\quad
     + w^*_c \pmatrix{0            & 0           & z_{13}     \cr
                      0            & z_{22}      & 0          \cr
                      z_{13}       & 0           & 0          \cr}; \cr
 }
 \eqn\MUP
$$
$ii)$ down-quark mass matrix
$$
 \eqalign{
  M_d &= v_b \pmatrix{-\c^2 y_{11} & 0           & \c y_{13}  \cr
                      0            & 0           & 0          \cr
                      -\c y_{13}   & 0           & y_{33}     \cr}
     + v^*_t \pmatrix{0            & \c^2 y_{12} & 0          \cr
                      \c^2 y_{12}  & 0           & -\c y_{23} \cr
                      0            & \c y_{23}   & 0          \cr} \cr
      &\quad
     + w^*_s \pmatrix{0            & 0           & z_{13}     \cr
                      0            & z_{22}      & 0          \cr
                      z_{13}       & 0           & 0          \cr}; \cr
 }
 \eqn\MDOWN
$$
$iii)$ charged-lepton mass matrix
$$
 \eqalign{
  M_e &= v_b \pmatrix{-\c^2 y_{11} & 0           & \c y_{13}  \cr
                      0            & 0           & 0          \cr
                      -\c y_{13}   & 0           & y_{33}     \cr}
     + v^*_t \pmatrix{0            & \c^2 y_{12} & 0          \cr
                      \c^2 y_{12}  & 0           & -\c y_{23} \cr
                      0            & \c y_{23}   & 0          \cr} \cr
      &\quad
    -3 w^*_s \pmatrix{0            & 0           & z_{13}     \cr
                      0            & z_{22}      & 0          \cr
                      z_{13}       & 0           & 0          \cr}. \cr
 }
 \eqn\MLEPT
$$
These mass matrices possess a desired hierarchical structure
within the coupling constants $y$'s and $z$'s of order one
provided $\c \ll 1$.

In the rest of this section, we investigate approximate relations
among masses of quarks and leptons implied by the above matrices.
Taking into account phenomenological hierarchy of masses
and smallness of Cabbibo-Kobayashi-Maskawa mixing angles,
we require $|v_t| \gg |v_b|$ and $|w_c| \gg |w_s|$,
which lead to the following
approximation to the mass matrices \MUP-\MLEPT:
\nextline
$i)$ up-quark mass matrix
$$
  M_u \simeq v_t \pmatrix{-\c^2 y_{11} & 0           & \c y_{13} \cr
                          0            & 0           & 0         \cr
                          -\c y_{13}   & 0           & y_{33}    \cr}
         + w^*_c \pmatrix{0            & 0           & z_{13}    \cr
                          0            & z_{22}      & 0         \cr
                          z_{13}       & 0           & 0         \cr};
$$
$ii)$ down-quark mass matrix
$$
  M_d \simeq v_b \pmatrix{0            & 0           & 0         \cr
                          0            & 0           & 0         \cr
                          0            & 0           & y_{33}    \cr}
         + v^*_t \pmatrix{0            & \c^2 y_{12} & 0         \cr
                          \c^2 y_{12}  & 0           & -\c y_{23}\cr
                          0            & \c y_{23}   & 0         \cr}
         + w^*_s \pmatrix{0            & 0           & 0         \cr
                          0            & z_{22}      & 0         \cr
                          0            & 0           & 0         \cr};
$$
$iii)$ charged-lepton mass matrix
$$
  M_e \simeq v_b \pmatrix{0            & 0           & 0         \cr
                          0            & 0           & 0         \cr
                          0            & 0           & y_{33}    \cr}
         + v^*_t \pmatrix{0            & \c^2 y_{12} & 0         \cr
                          \c^2 y_{12}  & 0           & -\c y_{23}\cr
                          0            & \c y_{23}   & 0         \cr}
        -3 w^*_s \pmatrix{0            & 0           & 0         \cr
                          0            & z_{22}      & 0         \cr
                          0            & 0           & 0         \cr}.
$$
These expressions of mass matrices resemble
the ansatz considered by Giudice
\refmark{\Giu}
in the context of supersymmetric grand unified theories.
In view of their simpler forms,
it is immediate to obtain approximate relations
$$
  m_b \simeq m_\t, \quad 3m_s \simeq m_\m, \quad m_d \simeq 3m_e,
  \quad \tan^2 \h_C \simeq {m_d \over m_s},
 \eqn\GJ
$$
where $\h_C$ is the Cabbibo angle.
These are the Georgi-Jarlskog relations\rlap,
\refmark{\Geo, \Bur, \Wil}
which are thought to be acceptable
at the unification scale.

%
{\caps 6. Renormalization-Group Analysis}

In the proceeding sections, we are mainly concerned with
the texture at and above the unification scale.
In order to compare the resulting masses and mixings with
experimental values, it is necessary to obtain
running couplings in low-energy region
by means of
renormalization-group (RG) equations.

We see running of gauge couplings
\refmark{\Des}
in this paper,
postponing the consideration on running of Yukawa couplings
to a subsequent paper.
Our postulate is that the light Higgs bosons
in the theory which do not decouple below the unification scale
consist of
$(2, 2, 1)_{\bf 10}$,
$(2, 2, 15)_{\o {\bf 126}}$, and $(1, 3, {\o {10}})_{\o {\bf 126}}\,$:
The standard Higgs doublet is given by a linear combination
of the four doublets contained in
$(2, 2, 1)_{\bf 10}$ and
$(2, 2, 15)_{\o {\bf 126}}$,
which has a mass of order $M_W$\rlap.
\refmark{\Del}
The remaining light Higgs bosons have masses of order $M_I$.

Then the one-loop RG running of gauge couplings are determined
from the present experimental values
\refmark{\Par}
below the weak scale
in the standard model (See Appendix A.2).
This implies that $M_I \sim 10^{11}$ GeV and $M_U \sim 10^{17}$ GeV.
Thus $\c = M_U/M_P$ may be of order $10^{-2}$, which seems appropriate
for the hierarchical mass matrices \MUP-\MLEPT\ to predict realistic
quark-lepton masses.
Note that $M_I$ turned out to be heavy enough to suppress
flavor-changing neutral current mediated by the Higgs bosons
with masses of order $M_I$.
It is also large enough to produce tiny masses for left-handed neutrinos
\refmark{\Fuk}
below experimental bounds\rlap.
\refmark{\Par}

%
{\caps 7. Conclusion}

We have presented the grand unified model in table 1
based on the symmetry group \U\ $\times$ \H.
Under the condition that the coupling constants
of order unity at the Planck scale
are arranged so as to produce necessary gauge hierarchy \SEQ,
we showed that the mass matrices naturally have
the hierarchical structure \MUP-\MLEPT\
and satisfy the Georgi-Jarlskog relations \GJ\
at the unification scale.
This is achieved by
the symmetry of the system with the
remnant effects of higher-dimensional Yukawa interactions
\FIVE-\SIX\ in spite of the minimality of the model.

Armed with the running gauge couplings,
we can make Yukawa couplings evolve from the unification
scale down to the weak scale.
We will investigate resultant
predictions of the model for
running masses and mixings in a separate paper.

%
\acknowledge

We would like to thank T.~Kugo and T.~Yanagida
for valuable discussions and encouragement.

%
\appendix
%

%
{\caps A.1. Mass Terms in \U\ Grand Unification}

In this appendix, we recapitulate explicit
contents of fermion bilinear terms in \U\ grand unified theory
\refmark{\Sak}
with left-handed fermions $\p$
in the representation {\bf 16}.
The group \U\ contains the Pati-Salam group
as a maximal subgroup:
$$
  {\rm SO}(6) \times {\rm SO}(4) \subset \U,
 \eqn\SUBS
$$
where ${\rm SO}(6) \simeq {\rm SU}(4)_{\rm C}$
and ${\rm SO}(4) \simeq {\rm SU}(2)_{\rm L} \times {\rm SU}(2)_{\rm R}$.
Let us denote an \U\ vector by
$$
  \f_j = (\f_A, \f_J); \quad
   A = 1, \cdots, 6;\ J = 7, 8, 9, 0,
 \eqn\VECT
$$
where $\f_A$ and $\f_J$ denote
SO$(6)$ and SO$(4)$ vectors, respectively.

Color-singlet neutral (CN) components in
the representation $(2, 2, 1)_{\bf 10}$
are given by
$$
  \VEV{\f_9} = v_1, \quad \VEV{\f_0} = v_2,
 \eqn\TENV
$$
which lead to a mass term
$$
  \p \p \VEV{\f_j} =
   (v_1 - iv_2) ({\bar u}_{_R} u_{_L} + {\bar \n}_{_R} \n_{_L})
    - (v_1 + iv_2) ({\bar d}_{_R} d_{_L} + {\bar e}_{_R} e_{_L}).
 \eqn\TENM
$$
In the case of $\VEV{H}$,
we write $v_t = v_1 - iv_2$ and $v_b = - v_1 - iv_2$.

The representation {\bf 120} decomposes under the Pati-Salam
group as follows:
$$
  {\bf 120} = (2, 2, 1) + (1, 1, 10) + (1, 1, {\o {10}})
             + (3, 1, 6) + (1, 3, 6) + (2, 2, 15).
 \eqn\DECTW
$$
The CN components in the representation $(2, 2, 1)_{\bf 120}$
are given by
$$
  \VEV{\f_{789}} = w_1, \quad \VEV{\f_{780}} = w_2,
 \eqn\TWV
$$
which yield a mass term
$$
  \p \p \VEV{\f_{jkl}} =
   (iw_1 + w_2) ({\bar u}_{_R} u_{_L} + {\bar \n}_{_R} \n_{_L})
    + (iw_1 - w_2) ({\bar d}_{_R} d_{_L} + {\bar e}_{_R} e_{_L});
 \eqn\TWM
$$
and those in the representation $(2, 2, 15)_{\bf 120}$ by
$$
 \eqalign{
  &\VEV{\f_{129}} = \VEV{\f_{349}}
                    = \VEV{\f_{569}} = v_1, \cr
  &\VEV{\f_{120}} = \VEV{\f_{340}}
                    = \VEV{\f_{560}} = v_2, \cr
 }
 \eqn\TWCV
$$
which result in a term
$$
  \p \p \VEV{\f_{jkl}} =
   -(iv_1 + v_2) ({\bar u}_{_R} u_{_L} - 3\,{\bar \n}_{_R} \n_{_L})
    + (iv_1 - v_2) ({\bar d}_{_R} d_{_L} - 3\,{\bar e}_{_R} e_{_L}).
 \eqn\TWCM
$$
Note that the fermion bilinear combinations in \TWM\ and \TWCM\
are antisymmetric in generation indices, as stated just below \COMP.

The CN components in the representation $(2, 2, 15)_{\o {\bf 126}}$
are characterized by
$$
 \eqalign{
  &\VEV{\f^*_{12789}} = \VEV{\f^*_{34789}}
                        = \VEV{\f^*_{56789}} = w^*_1, \cr
  &\VEV{\f^*_{12780}} = \VEV{\f^*_{34780}}
                        = \VEV{\f^*_{56780}} = w^*_2, \cr
 }
 \eqn\SIXV
$$
due to self-duality of the representation {\bf 126},
which provide a mass term
$$
  \p \p \VEV{\f^*_{jklmn}} =
   (w^*_1 - iw^*_2) ({\bar u}_{_R} u_{_L} - 3\,{\bar \n}_{_R} \n_{_L})
    + (w^*_1 + iw^*_2) ({\bar d}_{_R} d_{_L} - 3\,{\bar e}_{_R} e_{_L}).
 \eqn\SIXM
$$
In the case of $\VEV{\S^*}$,
we write $w^*_c = w^*_1 - iw^*_2$ and $w^*_s = w^*_1 + iw^*_2$.

Let us turn to the consideration on remnant effects of
higher-dimensional terms.
As an example, we investigate dimension-five terms
$\p \p H_j \F_{jklm}$ and $\p \p H_j \F_{klmn}$
of the form $\p \p H \F$.

The $(1, 1, 1)_{\bf 210}$ component in the field $\F_{jklm}$
is given by $\F_{7890}$, which has the VEV $\VEV{\F_{7890}} = M_U$.
Thus the term $\p \p H_j \F_{jklm}$ effectively act as a relevant term
$\p \p H_J \VEV{\F_{JKLM}}$
(suppressed by a factor $1/M_P$,
where $M_P$ is regarded as the cut-off scale),
which we call a remnant of the term $\p \p H_j \F_{jklm}$.

The field $H_J \VEV{\F_{JKLM}}/M_P$
behaves as a Higgs component in the representation {\bf 120}
which possesses VEVs
$$
  w_1 = {1 \over M_P} \VEV{H_0} \VEV{\F_{0789}}
      = - v_2 {M_U \over M_P}, \quad
  w_2 = {1 \over M_P} \VEV{H_9} \VEV{\F_{9780}} = v_1 {M_U \over M_P},
 \eqn\FVEV
$$
where $v_1, v_2$ and $w_1, w_2$ correspond to those defined by
\TENV\ and \TWV, respectively.
By virtue of \TWM, we obtain
$$
  \quad {1 \over M_P} \p \p \VEV{H_J} \VEV{\F_{JKLM}}
   = \c v_t ({\bar u}_{_R} u_{_L} + {\bar \n}_{_R} \n_{_L})
    + \c v_b ({\bar d}_{_R} d_{_L} + {\bar e}_{_R} e_{_L}),
 \eqn\IMAS
$$
where $\c = M_U/M_P$.

On the other hand,
the term $\p \p H_j \F_{klmn}$ yields a remnant
$\p \p H_A \VEV{\F_{KLMN}}$
with $A$ as an SO(6) index given in \VECT.
The color-triplet Higgs fields $H_A$ decouple
from the system below the unification scale (See section 6).
Hence
this remnant does not alter low-energy physics effectively,
to say nothing of
the mass matrices in the standard-model framework.

%
{\caps A.2. RG Equations for Gauge Couplings}

This appendix provides one-loop RG equations for gauge couplings
\refmark{\Sla}
in the setting described in section 6.

The one-loop RG equation for a gauge coupling
is generally given by
$$
  {d\w_i \over dt} = -{b_i \over 2\pi};
   \qquad \w_i = {4\pi \over g_i^2}, \quad t = \ln \m,
 \eqn\RGE
$$
where $g_i$ is a gauge coupling, $\m$ denotes a renormalization point,
and $b_i$ is a model-dependent constant.

Between the scales $M_W$ and $M_I$, the relevant gauge group
is the standard-model one $\SM$,
whose running gauge couplings evolve according to the constants
$$
  (b_i) = ({41 \over 10}, -{19 \over 6}, -7);
   \quad i = 1_{\rm Y}, 2_{\rm L}, 3_{\rm C}.
 \eqn\SMC
$$
Between the scales $M_I$ and $M_U$, the Pati-Salam group $\PS$
is the relevant one, and its gauge couplings evolve according to
the constants
$$
  (b_i) = (2, {26 \over 3}, -{7 \over 3});
   \quad i = 2'_{\rm L}, 2_{\rm R}, 4_{\rm C}.
 \eqn\PSC
$$

The boundary conditions are given by
$$
  \w_{1_{\rm Y}} = {3 \over 5}\w_{2_{\rm R}}+{2 \over 5}\w_{4_{\rm C}},
   \quad \w_{2_{\rm L}} = \w_{2'_{\rm L}},
    \quad \w_{3_{\rm C}} = \w_{4_{\rm C}}
 \eqn\BCSM
$$
at the scale $M_I$ and
$$
  \w_{2'_{\rm L}} = \w_{2_{\rm R}} = \w_{4_{\rm C}}
 \eqn\BCPS
$$
at the scale $M_U$,
where we have ignored threshold corrections.


\refout

\bye